\documentclass[preprint,showpacs,aps]{revtex4}

\usepackage{graphicx}% Include figure files
\usepackage{dcolumn}% Align table columns on decimal point
\usepackage{bm}% bold math

\begin{document}

\def\bbox#1{\hbox{\boldmath${#1}$}}
\def\gtsim{$\raisebox{0.6ex}{$>$}\!\!\!\!\!\raisebox{-0.6ex}{$\sim$}\,\,$}
\def\ltsim{$\raisebox{0.6ex}{$<$}\!\!\!\!\!\raisebox{-0.6ex}{$\sim$}\,\,$}
\def\pt{\bbox{p}_t}
\def\xx{\hbox{\boldmath{$ x $}}}
\def\pp{\hbox{\boldmath{$ p $}}}
\def\pt{\pp_{{}_T}}
\def\yb{{\bar y }}
\def\mt{m_{{}_T}}
\def\zb{{\bar z }}
\def\tb{{\bar t }}
\def\rhoe{ {\rho_{\rm egg}} }
\def\qq{\hbox{\boldmath{$ q $}}}

\title {\null\vspace*{-.0cm}\hfill { \small nucl-th/0302053 } \\
\vskip 0.8cm Intensity Interferometry for a Chaotic Source\\
with a Collective Flow and Multiple Scattering }

\author{Cheuk-Yin Wong}
\email{wongc@ornl.gov}
\affiliation{Physics Division, Oak Ridge National Laboratory, Oak Ridge,
TN 37831 USA}
\received{\today}

\begin{abstract}
We study the effects of a collective flow and multiple scattering on
two-particle correlation measurements in Hanbury-Brown-Twiss intensity
interferometry.  We find that under a collective flow the effective
source distribution in a two-particle correlation measurement depends
on the initial source distribution.  In addition, it depends on a
collective flow phase function which consists of terms that tend to
cancel each other.  As the detected particles traverse from the source
point to the freeze-out point, they are subject to multiple scattering
with medium particles.  We examine the effects of multiple scattering
on HBT correlations.  By using the Glauber theory of multiple
scattering at high energies and the optical model at intermediate
energies, we find that multiple scattering leads to an absorption and
an effective density distribution that depends on the initial source
distribution.

\end{abstract}


\maketitle

\section{Introduction}

Intensity interferometry, as first proposed by Hanbury-Brown-Twiss
(HBT) to measure the angular diameter of a star using the correlation
between two photons \cite{Hbt54}, has been applied to optical
coherence, subatomic physics, and heavy-ion collisions, utilizing
different types of particles \cite{Won94}-\cite{Mol02a}. Much
information on the space-time distribution of the emitting source is
obtained from these measurements.

Recent experimental measurements of HBT correlations in relativistic
heavy-ion collisions show only relatively small changes of the
extracted longitudinal and transverse radii as a function of collision
energies \cite{Phe02,Sta01,Lis00,Ahl02,Ahl00,Agg00}.  The small
variation of the HBT radii can be explained partly as due to the
space-time ordering of the momenta (correlation between the position
and the momentum) of the produced particles which was used to explain
similar weak collision-energy dependencies of HBT radii in $e^+e^-$,
$pp$, $p\bar p$ collisions (see, for example, Section 17.4 of Ref.\
\cite{Won94}).  In addition to space-time ordering of momenta,
produced particles are subject to collective flows.  It is important
to find out what physical quantities are measured by the two-particle
intensity interferometry for a chaotic source with a collective flow.
This will help us understand how collective flows may affect the size
parameters extracted in two-particle correlation measurements.  

As the detected particles traverse from the source point to the
freeze-out point, they are subject to final state mean-field
interactions and multiple collisions with particles in the medium.  The
effect of final mean-field potential has been studied earlier by
Gyulassy $et~al.$ \cite{Gyu79}.  They found that for a coherent source
the only effect of the mean field potential is to redistribute the
momentum distribution and the two-particle correlation function is not
affected.  For a chaotic source, the effect of a mean field is given
in terms of distorted wave functions in the mean field
\cite{Gyu79}.  The effects of the mean field in a chaotic source 
was further studied by Chu $et~al.$ \cite{Chu94} and Shoppa $et~al.$
\cite{Sho00}.  They found that for low-energy detected particles 
the Coulomb mean-field leads to substantial modification of the
two-particle correlation, but the effect is small for high-energy
detected particles.

The collisions of the detected particles with medium particles has
been considered to be a source of chaoticity.  It is usually assumed
that as a result of the random collisions of medium particles, the
initial source will evolve into a chaotic source at freeze-out.  The
source that is observed in HBT measurements will be the chaotic
freeze-out source, and the HBT radii will correspond to those of the
freeze-out configuration.  For example, in the emission function
modeling of \cite{Wie98,Wie99}, an analytical form of the freeze-out
phase-space distribution with a collective flow is assumed, and the
single-particle distribution and two-particle correlations are
described by a freeze-out distribution. In other examples,
hydrodynamics or covariant transport theory is followed until
freeze-out and the radii of the freeze-out configuration are then
extracted to compare with experimental HBT radii
\cite{Ris96,Hei02,Zsc02,Mol02,Mol02a}.  In such an analysis, there is
the outstanding puzzle that experimental measurements give $R_{\rm
out}/R_{\rm side} \approx 0.9-1.1$ while hydrodynamical predictions
yield $R_{\rm out}/R_{\rm side}$ significantly larger than 1.0 at
freeze-out \cite{Ris96,Hei02,Zsc02}.  The covariant transport theory
can explain the magnitude of $R_{\rm out}$ by assuming a large
opacity, but there remains the puzzle that $R_{\rm side}$ is
under-estimated \cite{Mol02,Mol02a}.
 
In the context of intensity interferometry, the question whether the
freeze-out configuration in a collective flow is a chaotic source
needs to be re-examined carefully as the problem of multiple
scattering must be treated properly.  Because the Hanbury-Brown-Twiss
intensity interferometry is purely a quantum-mechanical phenomenon,
the problem of multiple scattering must be investigated within a
quantum-mechanical framework.  We shall focus our attention to
high- and intermediate-energy detected particles that have
sufficient energies to propagate from the production point to the
detection point.  It is necessary to study the interference of waves
using the probability amplitudes in the multiple scattering process,
instead of the conventional description of incoherent collisions in
terms of probabilities and cross sections.  The Glauber theory of
multiple scattering \cite{Gla59} has been shown to be a valid
description for the interaction of a pion with the nuclear medium at a
pion energy from 300 to 1200 MeV \cite{Ari91,Ose91,Joh92}.  At lower energies, the optical model
\cite{Joh92,Sat92,Joh96,Hon99,Che93,Che95} has been found to give a
very good description of the interaction of a pion with a nucleus from
120 MeV to 766 MeV \cite{Hon99}.  They can be applied here to describe
the probability amplitudes for the propagation of energetic detected
particles (pions from 0.1 to 1 GeV, say). By using the Glauber theory
of multiple scattering at high energies and the optical model at
intermediate energies, we find that the multiple scattering process
leads to an absorption of particles and an HBT effective density
distribution that depends on the initial source distribution rather
than the freeze-out density distribution.

Previously, the intensity interference was studied in the framework of
probability amplitudes for the propagation of produced particles
\cite{Won94}.  We shall used the same framework to investigate here
the effects of a collective flow and multiple scattering.  We examine
first the single-particle distribution in Section II. The probability
amplitude for the propagation of a produced particle from the
production point to the detected point is obtained.  The sum over the
probability amplitudes from all the source points then leads to the
single-particle distribution which depends on the initial source
distribution.  In Section III, we study the two-particle momentum
distribution.  For a chaotic source, the quantum effect of
symmetrizing the amplitude for bosons (or anti-symmetrizing the
amplitude for fermions) with respect to the exchange of the two
particles leads to the phenomenon of intensity interferometry.  We
show that under a collective flow the two-particle correlation
function depends on the initial source distribution
and a collective flow phase function.  In Section IV, we study the
effects of multiple scattering of the detected particles with
particles in the medium.  In Section V, we discuss the relevance of
the present results to HBT measurements in high-energy heavy-ion
collisions.

\section{Single-Particle Distribution} 

Following arguments similar to those in Ref.\ \cite{Won94}, we
consider the production of a particle with four-momentum $\kappa=(\bbox
{\kappa},\kappa^0)$  at a source point $x=(\bbox{ x}, t)$ and its subsequent
detection with four-momentum $k=(\bbox{k},k^0)$ at the
space-time point $x_d=(\bbox{x}_d,t_d)$, as shown in Fig.\ 1.  For
convenience, we shall use the source center-of-mass system as the
reference frame to measure all momenta and space-time coordinates.

\begin{figure}[h]
\includegraphics{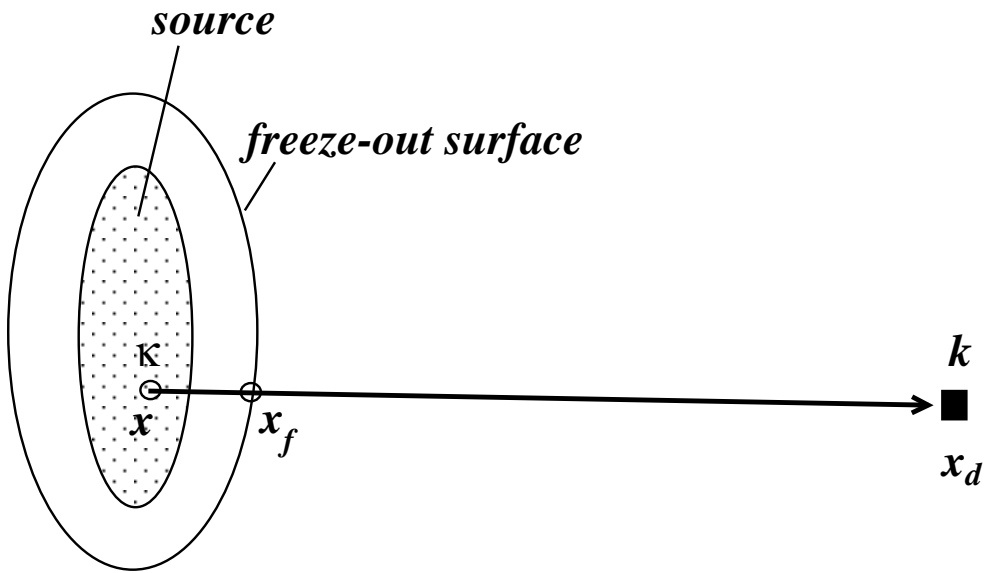}% Here is how to import EPS art
%\begin{center}
%\special{psfile=hbt1.eps hoffset=-30 voffset=-100 hscale=70
%  vscale=70 angle=0}
%\end{center}
\caption{ A particle of momentum $\kappa$ is emitted at a
typical source point $x$ of an extended source, is subject to a
collective flow to reach the freeze-out surface at $x_f$, and is
detected with momentum $k$ at the detection point $x_d$.  The straight
line joining $x$ to $x_d$ is the trajectory of the particle from $x$ to
$x_d$.  The figure is not drawn to scale.  The distance between $x$ and
$x_d$ is many orders of magnitude greater than the linear dimension of
the extended source.  }
\end{figure}

To describe the single-particle distribution and intensity
interference, we need the probability amplitude for the particle to be
produced at the source point $x$, to propagate to the freeze-out point
$x_f$, and to arrive at the detection point $x_d$.  We can
characterize the production probability amplitude for producing a
particle of momentum $\kappa$ at $x$ by a magnitude $A(\kappa x)$ and
a phase $\phi_0(x)$.  Without loss of generality, the functions
$A(\kappa x)$ and $\phi_0(x)$ can be taken to be real, and $A(\kappa
x)$ can be taken to be non-negative.  The magnitude $A(\kappa x)$ is
related to the initial phase-space distribution $f_{\rm init} (\kappa
(x),x)$ (see Eq.\ (\ref{ff1}) below).  The production phase
$\phi_0(x)$ describes the degree of coherence or chaoticity of the
particle production process. We shall be interested in an initially
chaotic source which can be represented by random and fluctuating
production phases $\phi_0(x)$ at the source points.

The complete probability amplitude $\Psi(\kappa  x \rightarrow k x_d)$
for a particle of momentum $\kappa $ to be produced from the source point
$x$, to propagate along the classical trajectory, and to arrive at
$x_d$ with momentum $k$ is
\begin{eqnarray} 
\Psi( \kappa   x \! \! \rightarrow \! k x_d) \! = \! A(\kappa x) e ^{i
\phi_0(x)} \psi( \kappa  x \to k x_d) \,,
\end{eqnarray}
where $\psi(\kappa  x \!\! \rightarrow \!\! k x_d)$ is the probability
amplitude for the propagation of the produced particle from the
production point $x$ to the freeze-out point $x_f$ and the detection
point $x_d$.  From the path-integral method, this amplitude is given by
\cite{Won94}
\begin{eqnarray}
\label{psi0}
\psi(\kappa x \to k x_d)= S({\rm classical~path,}~\kappa x \to k x_d)
= \exp \{-i\int_x^{x_d}
\kappa (x') \cdot dx'\},
\end{eqnarray}
where the integration is carried out along the classical trajectory,
and we have used the notation $\kappa (x')$ to represent the momentum
of the particle at the space-time point $x'=(\bbox{x}'t')$ with $t\le
t'\le t_d$. We consider a Lagrangian picture of collective motion in
which we follow the fluid element at $x'$ and its velocity field
$\bbox{\beta}(x')$.  For a given detected particle with momentum
$k=(k^0,\bbox{k})$, the initial momentum $\kappa $ depends on the
collective flow velocity $\bbox{\beta}(x_f)$ at freeze-out (see Eqs.\
(\ref{k1}), (\ref{k2}), and (\ref{rel1}) below).  The freeze-out
coordinate $x_f$ in turn depends on the initial source coordinate $x$.
Therefore, $\kappa(x) $ is a function of the initial source coordinate
$x$.

The momentum of the produced particle $\kappa (x')$ in different
space-time regions along the classical path from $x$ to $x_d$ is given
in the non-relativistic case by:
\begin{eqnarray}
\label{k1}
\bbox{\kappa}(x')=
\cases
{\bbox{\kappa}(x)+ m\bbox{\beta}(x'),
&~~{\rm for}~ $t\le t'\le t_f$, \cr
 \bbox{\kappa}(x)+ m\bbox{\beta}(x_f)= \bbox{k},        
&~~{\rm for}~ $t_f\le t'\le t_d$,\cr
}  
\end{eqnarray}
and
\begin{eqnarray}
\label{k2}
\kappa^0(x')=\cases
{
[\bbox{\kappa}(x)+m\bbox{\beta}(x')]^2 / 2m,
&~~~~~for~ $t\le t'\le t_f$, \cr 
[\bbox{\kappa}(x)+m\bbox{\beta}(x_f)]^2 / 2m=  k^0, 
&~~~~~for~ $t_f\le t'\le t_d. $\cr
} 
\end{eqnarray}
In the relativistic case, the momentum ${\kappa}(x')$ under a
collective flow with velocity $\bbox{\beta}(x')$ is given by
\begin{eqnarray}
\label{rel1}
\kappa(x')=\cases
{
\Lambda(\bbox{\beta}(x')) \kappa(x), 
&~~~~~for~ $t\le t'\le t_f$, \cr 
\Lambda(\bbox{\beta}(x_f))\kappa(x)=k,
&~~~~~for~ $t_f\le t'\le t_d, $\cr
} 
\end{eqnarray}
where $\Lambda(\bbox{\beta}(x'))$ is a matrix with elements
\begin{eqnarray}
\Lambda_0^0 ({\bbox{\beta}})&=&\gamma=1/\sqrt{1-|\bbox{\beta}|^2},
\nonumber\\
\Lambda_0^i ({\bbox{\beta}})&=&\Lambda_i^0 ({\beta}) =  \gamma \beta^i,
\\
\Lambda_k^i ({\bbox{\beta}})&=&\delta_k^i+{\gamma -1 \over \beta^2}
\beta^i \beta^k,
\nonumber
\end{eqnarray}
and $i,k=1,2,3$.  

The momentum of the particle does not change after freezing out at
$t'=t_f$.  Thus, the exponential function in $\psi(\kappa x \to k x_d)$
can be separated into the contribution from $x$ to $x_f$ and from
$x_f$ to $x_d$,
\begin{eqnarray}
\label{psi}
\psi(\kappa x\to k x_d)= \exp\{
- i\int_x^{x_f}\kappa (x')\cdot dx'
- i {k} \cdot (x_d-x_f) 
\}.
\end{eqnarray}

In addition the production from the source point $x$, the particle can
also be produced from other source points in the extended source.  The
total amplitude for the detection of a particle at $x_d$ is the sum of
the probability amplitudes from all source points.  After taking into
account the production probability amplitude $Ae^{i \phi}$ at
different source points, the total probability amplitude for a
particle with momentum $k$ to be produced from the extended source and
to arrive at the detection point $x_d$ is given by
\begin{eqnarray}
\Psi(k : \{ {}^{{\rm all}~x}_{\rm points} \} \rightarrow x_d) 
&=&
\! \sum_{x} A(\kappa (x),x) e ^{i \phi_0(x)} \psi(\kappa  x \! \rightarrow k x_d)
\nonumber \\
&=&
\! \sum_{x} A(\kappa (x),x) e ^{i \phi_0(x)}\, \exp\{
-i\int_x^{x_f} \kappa (x')\cdot dx' 
-i {k} \cdot (x_d-x_f) 
\} \,.       \end{eqnarray}

The single-particle momentum distribution, $P(k)$, which is the
probability for a particle of momentum $\kappa $ to be produced from the
extended source and to arrive at the detection point $x_d$ with
momentum $k$, is the absolute square of the total probability
amplitude,
\begin{eqnarray}
\label{pkk}
P(k) &=& 
|\Psi(k : \{ {}^{{\rm all}~x}_{\rm points} \} \rightarrow x_d) |^2
\nonumber \\
&=&
|\! \sum_{x} A(\kappa (x),x) e ^{i \phi_0(x)}\, 
\exp\{
-i\int_x^{x_f} \kappa (x')\cdot dx' 
-i  {k} \cdot (x_d-x_f) 
\} |^2 \,.        \end{eqnarray}
We expand the righthand side of Eq.\ (\ref{pkk}) into terms
independent of $\phi_0(x)$ and terms containing $\phi_0(x)$.  We obtain
\begin{eqnarray}
\label{pk}
P(k )= \sum_{x} A^2(\kappa (x),x)+ \sum_{ {x, y} \atop {x\ne y} }& &
A(\kappa (x),x) A(\kappa (y),y) e ^{i \phi_0(x)} e ^{-i \phi_0(y)} e^{
-i {k} \cdot (x_d-x_f) +i k \cdot (y_d-y_f)} \nonumber \\ & &
\times \exp\{- i\int_{x}^{x_f}\kappa (x') 
\cdot dx' +i\int_{y}^{y_f} \kappa (y')dy'
\} \,.
\end{eqnarray}   
For an initially chaotic source, the phases at different source points
can be described by a random and fluctuating phase $\phi_0(x)$ with a
period much shorter than the mean-life of the source.  For such a
source, the second term on the righthand side of the above equation
gives a zero contribution because the large number of terms with
slowly varying magnitudes, but rapidly fluctuating random phases,
cancel out one another in the sum over the source time coordinate (see
Supplement 17.1 of Ref.\ \cite{Won94}). The properties and the
validity of using the time average for a chaotic source have been
discussed in detail in Chapter X of the text of Born and Wolf
\cite{Bor64}.  Therefore, Eq.\ (\ref{pk}) becomes
\begin{eqnarray}
\label{pk1}
P(k) = \sum_{x} A^2(\kappa (x),x) \,. \end{eqnarray} The summation over
the source points necessitates the specification of the initial
density $\rho(x)$ of the source points per unit space-time volume at
the point $x$.  With this specification, the summation should be
transcribed as an integral over $x$,
\begin{eqnarray}
\label{tran}
\sum_x ... ~~~\rightarrow \int d^4 x~ \rho(x) ...  \end{eqnarray}
Therefore, we can rewrite Eq.\ (\ref{pk1}) as
\begin{eqnarray}
\label{pk2}
P(k) = \int d^4x \, \rho(x) A^2(\kappa (x),x) \,,
\end{eqnarray}
which is independent of $x_d$. We can compare this equation with the
properties of the phase space distribution function.  One can describe
the observed particles as coming from the initial phase space
distribution $f_{\rm init} (\kappa (x),x)$ defined by
\begin{eqnarray}
P(k) 
= \int d^4x \, f_{\rm init}(\kappa (x),x).
\end{eqnarray} 
A comparison of the above equation with Eq.\ (\ref{pk2}) shows that
the observed particle with momentum $k$ comes from the initial phase
space distribution
\begin{eqnarray}
\label{ff1} 
f_{\rm init}(\kappa (x),x) =
\rho(x) A^2(\kappa (x),x)  \,. 
\end{eqnarray}
One can alternatively describe the observed particles as coming from
the freeze-out phase space distribution $f_f (k, x_f)$ defined by
\begin{eqnarray}P(k) 
= \int d^4x_f \, f_f(k,x_f) = \int d^4x \, f_{\rm init}(\kappa (x),x).
\end{eqnarray} 
The initial distribution and the freeze-out distribution are
equivalent descriptions and they are related to each other by
\begin{eqnarray}
\label{ff}
f_f(k,x_f)= 
\left | {\partial (\bbox{x},     x^0) 
  \over  \partial (\bbox{x}_f,   x_f^0  )} \right |
f_{\rm init} (\kappa (x),x)
=
\left | {\partial (\bbox{x},   x^0) 
  \over  \partial (\bbox{x}_f, x_f^0  )} \right | 
\rho(x) A^2(\kappa (x),x)  \,,
\end{eqnarray}
where $\left | {\partial (\bbox{x}, x^0) / \partial (\bbox{x}_f, x_f^0
)} \right |$ is the Jacobian determinant arising from the mapping of
the initial source point $\{\bbox{x},x^0\}$ to the freeze-out point
$\{\bbox{x}_f,x_f^0\}$ due to the collective flow.

\section{Two-particle Distribution}

We consider the case in which a particle of momentum $\kappa_1(x_1)$
starts from $x_1$, propagates to the freeze-out point $x_{f1}$, and
arrives at the detection point $x_{d1}$ with momentum $k_1$, and
another identical particle of momentum $\kappa_2(x_2)$ starts from
$x_2$, propagates to the freeze-out point $x_{f2}$, and arrives at
$x_{d2}$ with momentum $k_2$, as indicated by the solid lines in Fig.\
2.  The probability amplitude for the production of the particle with
momentum $\kappa_j(x_i)$ at $x_i$ is given by
$A(\kappa_j(x_i),x_i)e^{i\phi_0(x_i)}$.  Therefore, the probability
amplitude for the two particles to be produced at the source points,
to propagate from the source points to the freeze-out points, and to
arrive at the detection points is
\begin{eqnarray}
\label{ak1}
A(\kappa_1(x_1),x_1) e^{i\phi_0(x_1)} A(\kappa_2(x_2), x_2) e^{i\phi_0(x_2)} 
&\exp\{
-i\int_{x_1}^{x_{f1}}\kappa_1(x') \cdot dx'
-i  {k}_1 \cdot (x_{d1}-x_{f1})
\}
\nonumber\\ 
\times 
&\exp\{
-i\int_{x_2}^{x_{f2}}\kappa_2(x') \cdot dx'
-i  {k}_2 \cdot (x_{d2}-x_{f2}) 
\}  \,.  
\end{eqnarray}

\begin{figure}[h]
\includegraphics{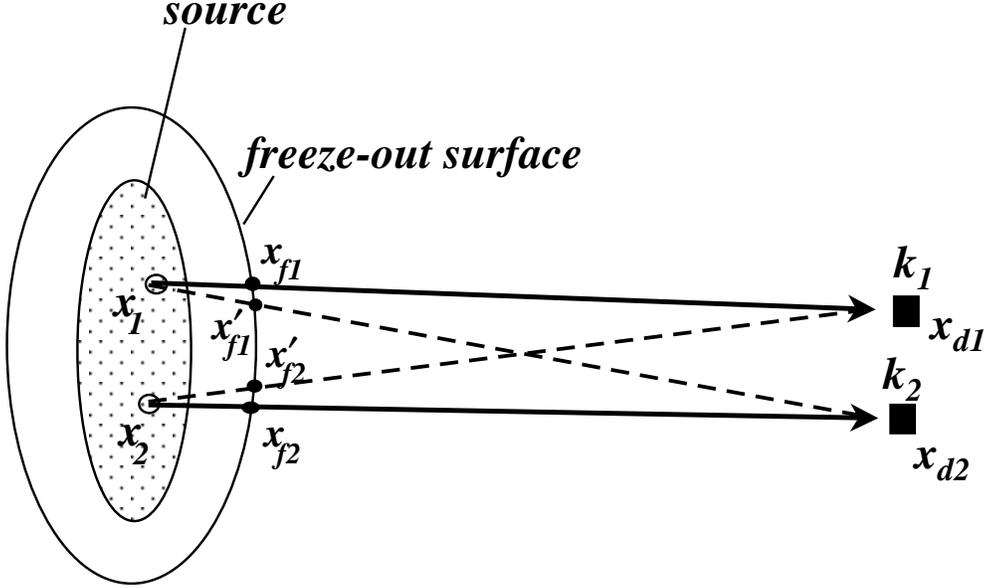}% Here is how to import EPS art
\caption{ A particle of momentum $k_1$ is detected at $x_{d1}$ and
another identical particle with momentum $k_2$ is detected at the
space-time point $x_{d2}$.  They are emitted from the source point
$x_1$ and $x_2$ of an extended source.  The solid lines joining $x_1
\to x_{f1} \to x_{d1}$ for $k_1$ and $x_2 \to x_{f2} \to x_{d2}$ for $k_2$, 
and the dashed lines joining $x_1 \to x_{f1}' \to x_{d2}$ for $k_2$
and $x_2 \to x_{f2}' \to x_{d1}$ for $k_1$ are possible trajectories.
}
\end{figure}

However, this is not the only probability amplitude contribution for
two identical particles produced from $x_1$ and $x_2$ to arrive at
$x_{d1}$ and $x_{d2}$. The particle of momentum $k_1$ detected at
$x_{d1}$ can also be produced at $x_2$ with momentum $\kappa_1(x_2)$ ,
propagate from $x_2$ to the freeze-out point $x_{f2}'(x_2)$, and
arrives at $x_{d1}$ with momentum $k_1$, while the other identical
particle of momentum $\kappa_2(x_1)$ starts at $x_1$, propagates from
$x_1$ to the freeze-out point $x_{f1}'(x_1)$, and arrives at $x_{d2}$
with momentum $k_2$, as indicated by the dashed lines in Fig.\ 2.
As the distances between the freeze-out points and the source points
are much smaller than the distances between the freeze-out points, and
the detection points and two-particle correlations occur for $k_1$
close to $k_2$, it is reasonable to make the approximations
$x_{f1}'=x_{f1}$ and $x_{f2}=x_{f2}'$ (see Fig.\ 2).  The probability
amplitude for this occurrence along the trajectories $x_1 \to x_{f1}'
\to x_{d2}$ for $k_2$ and $x_2
\to x_{f2}' \to x_{d1}$ for $k_1$ is
\begin{eqnarray}
\label{ak2}
A(\kappa_{1}(x_2), x_2) e^{i\phi_0(x_2)} A(\kappa_2(x_1),x_1) e^{i\phi_0(x_1)} 
&\exp\{
-i\int_{x_2}^{x_{f2}}\kappa_1(x') \cdot dx'
-i  {k}_1 \cdot (x_{d1}-x_{f2})
\}  
\nonumber\\ 
\times
& \exp\{
-i\int_{x_1}^{x_{f1}}\kappa_2(x') \cdot dx' 
-i  {k}_2 \cdot (x_{d2}-x_{f1})
\}
\,.  
\end{eqnarray}
Because of the indistinguishability of the particles and the
Bose-Einstein statistics of identical bosons (or the Fermi-Dirac
statistics for fermions), the probability amplitude must be
symmetrical (or antisymmetrical) with respect to the interchange of
the labels of the particles which distinguish them.  In this case, the
only labels which distinguish the two identical particles are the
source point coordinates $x_1$ and $x_2$, because one particle of
momentum $k_1$ has been determined to have been detected at $x_{d1}$ and
the other identical particle of momentum $k_2$ at $x_{d2}$.  The
probability amplitude must be symmetrical for bosons and
anti-symmetrical for fermions, with respect to the interchange of the
labels $x_1$ and $x_2$.  Accordingly, the probability amplitude which
satisfies this symmetry is the sum (or difference) of Eqs.\
(\ref{ak1}) and ({\ref{ak2}) divided by $\sqrt{2}$:
\begin{eqnarray} 
 {1\over \sqrt{2}}
\biggl \{  && A(\kappa_{1}(x_1), x_1) e^{i\phi_0(x_1)}
A(\kappa_{2}(x_2), x_2) e^{i\phi_0(x_2)} 
\exp\{
-i\int_{x_1}^{x_{f1}}\kappa_1(x') \cdot dx'
-i  {k}_1 \cdot (x_{d1}-x_{f1})
\}
\nonumber\\
&&~~~~~~~~~~~~~~~~~~~~~~~~~~~~~~~~~~~~~~~~~
\times
\exp\{
-i\int_{x_2}^{x_{f2}}\kappa_2(x') \cdot dx'
-i  {k}_2 \cdot (x_{d2}-x_{f2}) 
\} 
 \cr 
&& +\theta A(\kappa_{1}(x_2),x_2) e^{i\phi_0(x_2)}
A(\kappa_{2}(x_1), x_1) e^{i\phi_0(x_1)} 
\exp\{
-i\int_{x_2}^{x_{f2}}\kappa_1(x') \cdot dx'
-i  {k}_1 \cdot (x_{d1}-x_{f2})
\}  
\nonumber\\
&&~~~~~~~~~~~~~~~~~~~~~~~~~~~~~~~~~~~~~~~~~~
\times 
\exp\{
-i\int_{x_1}^{x_{f1}}\kappa_2(x') \cdot dx' 
-i  {k}_2 \cdot (x_{d2}-x_{f1})
\}
\} 
 \cr 
&&\equiv e ^{i
\phi_0(x_1)} e ^{i \phi_0(x_2)} 
\Phi (\kappa_1 \kappa_2, x_1 x_2 \rightarrow
x_{d1} x_{d2}) \,,  
\end{eqnarray} 
where $\theta$ is 1 for bosons, $-1$ for fermions, and $\Phi (\kappa_1
\kappa_2, x_1 x_2 \rightarrow x_{d1} x_{d2}) $ is the part of the probability
amplitude in the above equation which does not depend on $\phi_0$.  It
is defined by
\begin{eqnarray}
\label{PHI}
&&
\!\!\!\!\!\!\!\!\!\!\!\!\!\!\!\!\!\!\!\!
\Phi (\kappa_1 \kappa_2 : x_1 x_2 \rightarrow x_{d1} x_{d2})
\nonumber\\
&=& {1 \over \sqrt{2}}  
\biggl \{
A(\kappa_1(x_1), x_1) A(\kappa_2(x_2) x_2)
\exp\{
-i\int_{x_1}^{x_{f1}}\kappa_1(x') \cdot dx'
-i  {k}_1 \cdot (x_{d1}-x_{f1})
\}
\nonumber\\
&&~~~~~~~~~~~~~~~~~~~~~~~~~~~~~~~~~~~~
\times 
\exp\{
-i\int_{x_2}^{x_{f2}}\kappa_2(x') \cdot dx'
-i  {k}_2 \cdot (x_{d2}-x_{f2}) 
\} 
\nonumber\\
&&~~~~
+\theta
A(\kappa_1(x_2), x_2) A(\kappa_2(x_1), x_1)
\exp\{
-i\int_{x_2}^{x_{f2}}\kappa_1(x') \cdot dx'
-i  {k}_1 \cdot (x_{d1}-x_{f2})
\} 
\nonumber\\
&&~~~~~~~~~~~~~~~~~~~~~~~~~~~~~~~~~~~~~~~
\times 
\exp\{
-i\int_{x_1}^{x_{f1}}\kappa_2(x') \cdot dx' 
-i  {k}_2 \cdot (x_{d2}-x_{f1})
\}
\,.         
\end{eqnarray}
Besides originating from the source points $x_1$ and $x_2$, the two
particles can also be produced at other source points in the extended
source.  The total amplitude is the sum of amplitudes from all
combinations of two source points.  Therefore, the total probability
amplitude for two identical particles to be produced from two source
points in the extended source and to arrive at their respective
detection points $x_{d1}$ and $x_{d2}$ with momenta $k_1$ and $k_2$ is
\begin{eqnarray} 
\label{PSI}
\Psi(k_1 k_2 \!: \! \{  { {}^{{\rm all~}x_1x_2} _ {\rm points}}  
 \} \!
 \rightarrow  \! x_{d1} x_{d2}) \! = \!
\!\sum_{ \{x_1, x_2\}}
\! e ^{i \phi_0(x_1)} e ^{i \phi_0(x_2)} 
&[&\Phi(\kappa_1 \kappa_2 \! : \! x_1 x_2  \! \rightarrow \! x_{d1} x_{d2})
\nonumber\\ 
&+&\theta \Phi(\kappa_1 \kappa_2 \! : \! x_2 x_1  \! \rightarrow \! x_{d1} x_{d2}) ]
\end{eqnarray}
where the sum is carried over distinct combinations of $\{x_1,x_2\}$.
The two-particle momentum distribution $P(k_1, k_2)$ is defined as the
probability distribution for two particles of momenta $k_1$ and $k_2$
to be produced from the extended source and to arrive at their
respective detection points $x_{d1}$ and $x_{d2}$.  From Eq.\
(\ref{PSI}), it is given by
\begin{eqnarray}
\label{PK}
P(k_1, k_2) 
= {1 \over 2 !}
| \Psi(\kappa_1 \kappa_2 \!: \! \{  { {}^{{\rm all~}x_1x_2} _ {\rm points}}   \} 
 \rightarrow  \! x_{d1} x_{d2})|^2
\,. 
\end{eqnarray}
We now apply the results of Eqs.\ (\ref{PHI})-(\ref{PK}) to examine
the momentum correlations of a chaotic source. A chaotic source is
described by a phase function $\phi_0(x)$ that is a random and
fluctuating function of the source point coordinate $x$.  For a
chaotic source, we again make use of the random and fluctuating nature
of the phases by substituting Eq.\ (\ref{PHI}) into Eq.\ (\ref{PSI})
and expand the righthand side of Eq.\ (\ref{PK}). We separate out
terms which are independent of $\phi_0$ and terms which contain $\phi_0$.
We obtain
\begin{eqnarray}
P(k_1, k_2)  =  
{1\over 2}   \! \sum_{x_1, x_2}
\! \! \biggl \{
\Phi^*(\kappa_1 \kappa_2 \! : \! y_1 y_2  \!\! \rightarrow \!\! x_{d1} x_{d2})
\vert_{y_1=x_1 \atop  y_2=x_2} 
\Phi(\kappa_1 \kappa_2 \! : \! x_1 x_2  \!\! \rightarrow \!\! x_{d1} x_{d2})
~~~~~~~~~~~~~~~~~~~~~~~~~~~~~~~~~~~~~~~~~
\nonumber
\end{eqnarray}
\vskip -1.3cm
\begin{eqnarray}~~~~~~~~~~~~~~~~~~~
+\theta
\Phi^*(\kappa_1 \kappa_2 \! : \! y_1 y_2  \!\! \rightarrow \!\! x_{d1} x_{d2})
\vert_{y_2=x_1 \atop  y_1=x_2}   
\Phi(\kappa_1 \kappa_2 \! : \!  x_1 x_2  \!\! \rightarrow \!\! x_{d1} x_{d2})
\!\! \biggr \}
~~~~~~~~~~~~~~~~~~~~~~~~~~~~~~~~~~~~~~~~
\nonumber
\end{eqnarray}
\vskip -0.7cm
\nobreak
\begin{eqnarray}+ {1\over 2} \!\!\!\! \sum_{x_1 , x_2 , y_1 , y_2 \atop 
{
\{x_1x_2\} \ne  \{y_1y_2\} } }
\!\!\!\!
\biggl \{ e ^{i \phi_0(x_1) + i\phi_0(x_2)
-i \phi_0(y_1) -i \phi_0(y_2)} 
~~~~~~~~~~~~~~~~~~~~~~~~~~~~~~~~~~~~~~~~~
\nonumber
\end{eqnarray}
\vskip -1.4cm
\nobreak
\begin{eqnarray}
\label{PK1}
~~~~~~~~~~~~~~~~ \times
\Phi^*(\kappa_1 \kappa_2: y_1 y_2  \!\! \rightarrow \!\! x_{d1} x_{d2})
\Phi(\kappa_1 \kappa_2: x_1 x_2  \!\! \rightarrow \!\! x_{d1} x_{d2})
\biggr \}
\,.         
\end{eqnarray}
\vskip 0.8cm
The two terms in the first summation on the righthand side are equal
because of exchange symmetry.  For the chaotic source, the last term
in Eq.\ (\ref{PK1}) gives a zero sum because the contributions of a
large number of terms with similar magnitudes but random and
fluctuating phases cancel out.  Therefore, Eq.\ (\ref{PK1}) becomes
\begin{eqnarray}
\label{PK2}
P(k_1,k_2) = \sum_{x_1 , x_2 } | \Phi(\kappa_1 \kappa_2: x_1 x_2 \rightarrow
x_{d1} x_{d2})|^2\,,  
\end{eqnarray} 
Converting the summations in Eq.\ (\ref{PK2}) into integrals with the
transcription (\ref{tran}), we can rewrite the total probability as a
double integral over the source point coordinates $x_1$ and $x_2$,
\begin{eqnarray}
P(k_1,k_2)
= \! \int d^4x_1 d^4x_2~ \rho(x_1) \rho(x_2)
| \Phi(\kappa_1 \kappa_2: x_1 x_2 \rightarrow x_{d1} x_{d2})|^2 . 
\end{eqnarray}
In the above equation, $|\Phi|^2$ can be obtained by using Eq.\
(\ref{PHI}).  We get
\begin{eqnarray}
|\Phi|^2&=&A^2(\kappa_1(x_1), x_1) A^2(\kappa_2(x_2) x_2)
\nonumber\\
& &+\theta A(\kappa_1(x_1), x_1) A(\kappa_2(x_2) x_2)
A(\kappa_1(x_2), x_2) A(\kappa_2(x_1), x_1){e^{i({\cal A}+{\cal B})}
+ e^{-i({\cal A}+{\cal B})} \over 2},
\end{eqnarray}
where
\begin{eqnarray}
{\cal A}&=&-\int_{x_1}^{x_{f1}}\kappa_1(x')\cdot dx' 
           -\int_{x_2}^{x_{f2}}\kappa_2(x')\cdot dx' 
           +\int_{x_1}^{x_{f1}}\kappa_2(x')\cdot dx' 
           +\int_{x_2}^{x_{f2}}\kappa_1(x')\cdot dx' 
\nonumber\\
&=&
           -\int_{x_1}^{x_{f1}}[\kappa_1(x')-\kappa_2(x')]\cdot dx' 
           -\int_{x_2}^{x_{f2}}[\kappa_2(x')-\kappa_1(x')]\cdot dx' ,
\end{eqnarray}
and
\begin{eqnarray}
\label{34}
{\cal B}&=& k_1 \cdot (x_{f1}-x_{d1})
          + k_2 \cdot (x_{f2}-x_{d2})  
          - k_1 \cdot (x_{f2}-x_{d1}) 
          - k_2 \cdot (x_{f1}-x_{d2}) 
\nonumber\\
&=& (k_1 -k_2)\cdot x_{f1}
 +  (k_2 -k_1)\cdot x_{f2}.
\end{eqnarray}
Therefore, we have
\begin{eqnarray}
&& P(k_1,k_2) 
= \int d^4x_1 
\rho(x_1)
A^2 (\kappa_1(x_1), x_1) \int d^4x_2 \rho(x_2) A^2(\kappa_2(x_2), x_2)
~~~~~~~~~~~~~~~
\nonumber \\
&&+ \theta  \int  d^4x_1 \rho(x_1)
A(\kappa_1(x_1),  x_1) A(\kappa_2(x_1), x_1) 
e^{ i(k_1-k_2) \cdot x_1 }
\exp \{ -i \int_{x_1}^{x_{f1}}\{[\kappa_1(x')-\kappa_2(x')]-[k_1-k_2] \}\cdot dx' \}
\nonumber \\
&& \times
 \int  d^4x_2 \rho(x_2)
A(\kappa_2(x_2),  x_2) A(\kappa_1(x_2), x_2) 
e^{-i(k_1-k_2) \cdot x_2 }
\exp \{ -i \int_{x_2}^{x_{f2}}\{[\kappa_1(x')-\kappa_2(x')]-[k_1-k_2] \}\cdot dx' \}
\nonumber \\
\end{eqnarray}
Using Eq.\ (\ref{pk2}),
we can rewrite the above equation as
\begin{eqnarray}
\label{pk1k2}
&&
\!\!\!\!\!\!\!\!\!\
P(k_1,k_2) 
= P(k_1) P(k_2)
 \nonumber\\
&&\!\!\!\!\!\!\!\!+ \theta 
\biggl \vert \! \int d^4x e^{ i(k_1 -k_2) \cdot x} 
\exp \{ -i \int_{x}^{x_{f}}\{[\kappa_1(x')-\kappa_2(x')]-[k_1-k_2] \}
\cdot dx' \}
\rho(x) A(\kappa_1(x),  x) A(\kappa_2(x), x) 
\biggr \vert ^2,\nonumber\\  
\end{eqnarray}
which is independent of $x_d$.  It is convenient to 
 rewrite the two-particle
distribution function as
\begin{eqnarray}
\label{pk1k2d}
P(k_1,k_2) = P(k_1) P(k_2) \biggl ( 1+ \theta \biggl \vert \! \int
d^4x \, e^{ i(k_1-k_2) \cdot x +i\phi_c(x, k_1 k_2 )}
\rho_{\rm eff} (x; k_1 k_2) \biggr \vert ^2 \biggr ),
        \end{eqnarray}
where $\rho_{\rm eff}$ is the effective density 
\begin{eqnarray}
\label{flow}
\rho_{\rm eff} (x; k_1 k_2)=
{ \rho(x) A(\kappa_1(x),  x) A(\kappa_2(x), x) \over
\sqrt{ P(k_1) P(k_2) } } \,,
\end{eqnarray}
and $\phi_c(x, k_1 k_2)$ is the collective flow phase function
\begin{eqnarray}
\label{gkk}
\phi_c (x, k_1 k_2)=
 - \int_{x}^{x_{f}}
\{[\kappa_1(x')-\kappa_2(x')]-[k_1-k_2] \}\cdot dx' 
\,.
\end{eqnarray}
The HBT two-particle correlation function
$R(k_1,k_2)=P(k_1,k_2)/P(k_1)P(k_2)-1$ becomes
\begin{eqnarray}
\label{rkk}
R(k_1,k_2)
=  \theta \biggl \vert \! \int d^4x \, 
e^{ i(k_1-k_2) \cdot x + i \phi_c(x, k_1 k_2)}
\rho_{\rm eff}(x; k_1 k_2) 
\biggr \vert ^2 .
\end{eqnarray}
From Eq.\ (\ref{rel1}), we have
\begin{eqnarray}
[\kappa_1(x')-\kappa_2(x')]-[k_1-k_2]
=[\Lambda(\bbox{\beta}(x'))-\Lambda(\bbox{\beta}(x_f))]
\Lambda^{-1}(\bbox{\beta}(x_f)) (k_1-k_2).
\end{eqnarray}
The collective flow phase $\phi_c(x, k_1 k_2)$ becomes
\begin{eqnarray}
\label{phic}
\phi_c (x, k_1 k_2)=
 - \int_{x}^{x_{f}} dx' \cdot 
\left \{
[\Lambda(\bbox{\beta}(x'))-\Lambda(\bbox{\beta}(x_f))]
\Lambda^{-1}(\bbox{\beta}(x_f)) (k_1-k_2)
\right \}
\,.
\end{eqnarray}
Thus, under a collective flow the two-particle correlation function
depends on the initial source distribution and a
collective flow phase function $\phi_c (x, k_1 k_2)$, as given by
Eqs.\ (\ref{pk1k2d})-(\ref{rkk}).  When there is no
collective flow, $\phi_c(x,k_1 k_2)$ is zero, and $P(k_1,k_2)$ and
$R(k_1, k_2)$ reduce to the usual two-particle correlation functions of
Eqs.\ (17.29) and (17.38) in Ref.\ \cite{Won94}.

We note that the collective flow phase function $\phi_c(x, k_1 k_2)$
consists of the differences
$\Lambda(\bbox{\beta}(x'))-\Lambda(\bbox{\beta}(x_f))$ and $k_1-k_2$
which can lead to substantial cancellation.  In fact, if $\beta(x')$
reaches the asymptotic velocity $\bbox{\beta}(x_f)$, then there will
be no contribution from this path element $dx'$ at $x'$ to the phase
function $\phi_c(x, k_1 k_2)$.

The velocity profile $\bbox{\beta}(x')$ of a source can be obtained by
solving the relativistic hydrodynamical equations.  We can get some
ideas on the expansion scenario for a highly compressed relativistic
ideal fluid initial at rest from the results of Rischke and Gyulassy
\cite{Ris96}. We can consider a spherical fireball in
which the collective flow is along the radial direction. (The
transverse expansion of a Bjorken cylinder can be discussed in a
similar way \cite{Ris96}.) When this highly compressed fluid is
allowed to evolve hydrodynamically, fluid elements at the surface
reach their asymptotic velocity (close to the speed of light) very
rapidly --- nearly at the beginning of the expansion.  Fluid elements
at the central region remain essentially at rest until they are
carried outward by the strong rebounding wave, and they reach the
asymptotic velocity also rapidly.

From Fig.\ 2 of \cite{Ris96}, the velocity of the fluid element
initial at $(r,\theta,\phi)$ is approximately
\begin{eqnarray}
\bbox{\beta}(x')=
 \bbox{\beta}(x_f) ~\theta(t'-t_c) ~~~~~{\rm for~~} t\le t' \le t_f,
\end{eqnarray}
where $t_c=t+2(R-r)$, $r\le R$, and $R$ is the radius of the initial
source.  The trajectory for the fluid element is
$r'(t')=r+\beta(x')t'$, $\theta'(t')=\theta,$ and
$\phi'(t')=\phi$. For such a collective flow, we can describe this as
an initial source up to $t=t_c$, and the source undergoes collective
flow at $t > t_c$ with a flow velocity reaching the asymptotic
velocity approximately in a step-wise manner.  Fluid elements reaching
the asymptotic velocity do not contribute to the collective phase
$\phi_c(x,k_1 k_2)$, as one notes previously in Eq.\ (\ref{phic}).
The collective flow phase function $\phi_c(x, k_1 k_2)$ is
approximately zero for this case.  For other initial conditions and
compression, the radial expansion and rapid rise of the collective
velocity to the asymptotic velocity should be similar, as the
collective motion arises from the expansion of a compressed matter
into an empty vacuum.  The magnitude of the asymptotic velocity may
depend on the degree of compression (and therefore on the heavy-ion
collision energy).  The collective flow phase function $\phi_c$ will
likely remain small because of the substantial cancellation in Eq.\
(\ref{phic}) mentioned above. We should expect that the effective
density measured in HBT measurements should depend essentially on the
initial source distribution.  Much more work will be required to
examine further the behavior of $\phi_c(x, k_1 k_2)$ and $R(k_1 k_2)$
for general flows and initial conditions.

\section {Effects of Multiple Scattering and Optical Potential}

The observation of the intensity interferometry depends on the degree
of coherence or chaoticity of the source.  As the initial source
particles traverse from the source point $x_{i}$ to the freeze-out
point $x_{fi}$ under a collective flow, they are subject to multiple
scattering with medium particles.  One may think that as a result of
these scatterings, the source becomes chaotic at freeze-out and the
distribution observed in an HBT measurement should correspond to the
freeze-out distribution
\cite{Wie98,Wie99}.

The problem of multiple scattering must however be treated properly in
the context of intensity interferometry.  As the Hanbury-Brown-Twiss
effect is purely a quantum-mechanical effect, the problem of multiple
scattering must be investigated within the quantum-mechanical
framework in terms of probability amplitudes instead of incoherent
collisions involving probabilities and cross sections.

Within such a theoretical framework, we can study our problem in the
Glauber model \cite{Gla59,Ari91} or the optical model
\cite{Sat92,Joh96,Hon99,Joh92,Che93,Che95}, depending on the energy of the
detected particle relative to the medium.  The Glauber theory has been
applied successfully to study the interaction of a pion with the
nuclear medium at a pion energy from 300 to 1200 MeV
\cite{Ari91,Ose91,Joh92}.  It can therefore be used to study the
interaction of an energetic pion with the hadron medium produced in
high-energy heavy-ion collisions.  At these energies, the trajectory
can be approximately described as a straight line, the probability
amplitude for the particle to traverse from $x$ to $x'$ after
suffering multiple scatterings with particles in the medium is
\begin{eqnarray}
\label{chi}
\exp \{i\phi_s(x\to x') \}=\exp \{i\sum_{j=2}^N 
\chi({\bbox{x}}_\perp'  - {\bbox{s}_{j\perp}})
\theta({\bbox{x}}_{\|}'  - {\bbox{s}_{j\|}} )
\theta({\bbox{s}_{j\|}} - {\bbox{x}}_\|  ) \}
,
\end{eqnarray}
where the function $\chi({\bbox{x}}_\perp' - {\bbox{s}_{j\perp}})$ is
the Glauber phase shift function for the interaction of the produced
particle with a medium particle at $\bbox{s}_j$ and the sum $\sum_j$
is carried over all medium particles having coordinates
$x_j=(t_j,\bbox{s}_{j \|}, \bbox{s}_{j\perp}), j=2,3,...,N$
\cite{Gla59}.  The phase shift function $\chi({\bbox{b}})$ can be 
obtained from two-body scattering data and is an analytical function
of the transverse coordinate $\bbox{b}$.  The subscript $\perp$ in
Eq.\ (\ref{chi}) denotes the component transverse to the particle
trajectory and the subscript $\|$ denotes the component along the
trajectory.

The wave function Eq.\ (\ref{chi}) from Glauber multiple scattering
theory contains a wealth of relevant information.  It depends on the
coordinates of all the particles with which the incident particle has
interacted. It treats correctly the case of no scattering and multiple
scattering, even up to the extreme case of $N-1$ scatterings in
succession.  It makes no difference whether the medium particles are
dense in close proximity or dilute in far separation.  Information on
the density of medium particles can be provided when one integrates
out the distribution of the medium particles.

One can show that starting with the multiple scattering wave function
Eq.\ (\ref{chi}), one can construct the Wigner function in the
transverse and longitudinal degrees of freedom.  The multiple
scattering wave function Eq.\ (\ref{chi}) leads naturally to to the
diffractive transport theory, and the equation of motion for the
Wigner function is just the Boltzmann equation \cite{Won03b}.  The
solution of the Wigner function can be decomposed into
multiple-scattering components each of which corresponds to the
scattering of the incident particle with a definite number of medium
particles.  The transverse width of the $n$-scattering component
increases with $\sqrt{n}$ and the maximum of the longitudinal
distribution of the $n$-scattering component is located longitudinally
at $n$ times the mean-free path.

As Glauber multiple scattering theory gives the classical diffractive
transport theory on the one hand, and retains the wave nature of the
propagation on the other hand, it is therefore appropriate to use
Glauber multiple scattering wave function to investigate the effects
of multiple scattering in intensity interferometry.

Using this description for the multiple scattering of the particle,
the probability amplitude for the produced particle to travel from the
source point $x$ to $x_d$ becomes
\begin{eqnarray} 
\label{sx2}
\Psi( \kappa  x \! \! \rightarrow \! k x_d) \! = \! A(\kappa x) e ^{i
\phi(x)} 
\psi(\kappa x \to k x_d)  
\end{eqnarray}
where $\phi(x)$ is 
\begin{eqnarray}
\phi(x)=\phi_0(x)+\phi_s(x\to x_f),
\end{eqnarray}
$\phi_0(x)$ is the initial random and fluctuating phase due to the
chaoticity of the production source, and $\phi_s(x\to x_f)$ is
\begin{eqnarray}
\phi_s(x\to x_f)=\sum_{j=2}^N 
\chi({\bbox{x}}_{f\perp}  - {\bbox{s}_{j\perp}})
\theta({\bbox{x}}_{f\|}  - {\bbox{s}_{j\|}} )
\theta({\bbox{s}_{j\|}} - {\bbox{x}}_\|  ).
\end{eqnarray}
As $x_f$ is a function of $x$, the phase function $\phi_s(x\to x_f)$
is a function of $x$ and can be abbreviated as $\phi_s(x)$.  It
contains a real and imaginary part.  The imaginary part is given by
\begin{eqnarray}
{\cal I}{ m}~ \phi_s(x)=\sum_{j=2}^{N} 
{\cal I}{ m}~
\chi({\bbox{x}}_{f \perp}  - {\bbox{s}_{j\perp}})
\theta({\bbox{x}}_{f\|}  - {\bbox{s}_{j\|}} )
\theta({\bbox{s}_{j\|}} - {\bbox{x}}_\|  ),
\end{eqnarray}
and it represents the degree of absorption as the detected particle
passes through the medium.

With this modification, the derivation of the single-particle and
two-particle distributions can be carried out as in the last sections.
As the multiple scattering bring in a complex phase $\phi_s(x)$ that
depends on $x$, the quantum description of the scattering amplitude
leads to the single-particle distribution
\begin{eqnarray}
\label{pk1mscat}
P(k) = \int dx_2 dx_3 dx_4 ...  dx_N  
\sum_{x} e^{i(\phi_s(x)-\phi_s^*(x))} A^2(\kappa (x),x)
\rho_{\rm med} (x_2, x_3, x_4,..., x_N)
\,,
\end{eqnarray}
where $\rho_{\rm med} (x_2, x_3, x_4,..., x_N)$ is the distribution of
the medium particles.  Note that the real part of $\phi_s(x)$ cancel
out in the factor $\phi_s(x)-\phi_s^*(x)$ in the above equation.
Consequently, when the multiple scattering between the detected
particle and the medium is taken into account, the single-particle
distribution for a collective flow is
\begin{eqnarray}
\label{pk2mscat}
P(k) = \int d^4x \, 
e^{-2 ~{\cal I}{m}~ {\bar \phi}_s(x)} \rho(x) A^2(\kappa (x),x)  \,, 
\end{eqnarray}
where 
\begin{eqnarray}
e^{-2 ~{\cal I}{m}~ {\bar \phi}_s(x)}
= \int dx_2 dx_3 dx_4....dx_N  
e^{-2 {\cal I}m\, \phi_s(x)}
\rho_{\rm med} (x_2, x_3, x_4,..., x_N)
.
\end{eqnarray}
In this case, we have
\begin{eqnarray}
P(k) = \int d^4x \, 
f_{\rm init}(\kappa (x),x),
\end{eqnarray}
where
\begin{eqnarray}
\label{pk3mscat}
f_{\rm init}(\kappa (x),x)=
e^{-2 ~{\cal I}m~ {\bar \phi}_s(x)} 
\rho(x) A^2(\kappa (x),x)  \,. \end{eqnarray}
The two-particle distribution for a system with a collective flow can
be obtained as in the last section, and we get
\begin{eqnarray}
\label{2b1mscat}
P(k_1,k_2) 
&=& P(k_1) P(k_2)
\nonumber\\
&+& \theta
\biggl \vert \! \int d^4x e^{ i(k_1-k_2) \cdot x  + i \phi_c(x, k_1 k_2) } 
e^{-2 ~{\cal I}{m}~ {\bar \phi}_s(x)}
\rho(x)
A(\kappa_1(x),  x) A(\kappa_2(x), x)
\biggr \vert ^2,         
\end{eqnarray}
or
\begin{eqnarray}
\label{pk1k2dd}
P(k_1,k_2) = P(k_1) P(k_2) \biggl ( 1+ \theta \biggl \vert \! \int
d^4x \, e^{ i(k_1-k_2) \cdot x +i\phi_c(x, k_1 k_2 )}
e^{-2 ~{\cal I}{m}~ {\bar \phi}_s(x)}
\rho_{\rm eff} (x; k_1 k_2) \biggr \vert ^2 \biggr ),
\end{eqnarray}
where $\phi_c(x, k_1 k_2 )$ is given by Eq.\ (\ref{gkk}) and
$\rho_{\rm eff} (x; k_1 k_2)$ is given by Eq.\ (\ref{flow}).  Thus,
multiple scattering leads to an absorption of produced particles as
they traverse from the source point to the freeze-out point.  There is
an absorption factor in both the single-particle distribution and the
two-particle correlation function.  The effective density distribution
$\rho_{\rm eff}$ revealed by HBT two-particle correlation measurements
depends on the initial source distribution and not necessarily on the
freeze-out density distribution.

At lower energies, the interaction of the pion with the medium can be
described by an optical model
\cite{Sat92,Joh96,Hon99,Joh92,Che93,Che95}.  For example, in the
interaction of a pion with a nucleus, a local phenomenological optical
potential has been successfully applied to explain the $\pi$-nucleus
elastic scattering data from 120 MeV to 766 MeV
\cite{Sat92,Joh96,Hon99}.  The effects of intermediate resonances such
as $\Delta(1232)$ shows up as giving rise to a peak in the imaginary
part of the optical potential near the $\Delta(1323)$ resonance
energy.  The optical potential is normally non-local but can be
described in terms of a local-equivalent potential by redefining some
kinematic quantities \cite{Sat92}.  A ``model-exact' microscopic
description of the optical potential has also been developed to
describe the interaction of a pion with a nuclear medium in terms of
the interaction between the pion and medium particles, including the
effects of the $\Delta(1323)$, $\Delta_{13}(1520)$, and $F_{15}(1680)$
resonances
\cite{Che93,Che95}.

Based on the successes of the optical model in describing the
interaction between a pion and nuclear matter, it is reasonable to use
similar descriptions to study the dynamics of an intermediate-energy
pion in hadronic matter produced in high-energy heavy-ion collisions.
An optical model is specially appropriate as the Hanbury-Brown-Twiss
intensity interferometry arises from an optical interference of the
wave amplitudes.

One can accordingly introduce a complex optical potential $V(k,x)$
with a negative imaginary part to describe the interaction between the
detected particle and the medium particles as it travels from the
source point to the freeze-out point.  The optical potential $V$
depends on the momentum of the particle $k$.  The momentum dependence
is particularly pronounced in the neighborhood of a resonance. In
principle, the optical potential can be determined from two-body data
and the distribution of particles in the medium, as carried out for
example in Ref.\ \cite{Joh92,Che93,Che95}.  Under the interaction of
an optical potential $V(k,x)$, the amplitude for a produced particle
to propagate from $x$ to $x''$ is given by
\cite{Gla59}
\begin{eqnarray}
\exp \{i\phi_s(x\to x'')\}= \exp \{-i\int_{x}^{x''}  {1 \over v}
V(k,x') dx_\|'\},
\end{eqnarray}
where $v$ is the magnitude of the velocity of the detected particle
relative to the hadron medium along the direction of propagation.
Therefore, when the multiple scattering between $x$ and $x_f$ is taken
into account, the probability amplitude for the particle to propagate
from $x$ to $x_d$ becomes
\begin{eqnarray} 
\label{sx3}
\Psi( \kappa  x \! \! \rightarrow \! k x_d) \! = \! A(\kappa x) e ^{i
\phi(x)} 
\psi(\kappa x \to k x_d),
\end{eqnarray}
where 
\begin{eqnarray}
\label{vxx}
\phi(x)=\phi_0(x)-\int_{x}^{x_d}  {1 \over v}
V(k,x') dx_\|'.
\end{eqnarray}
The phase shift due to the optical potential, $\phi_s(k,x)$, is a
complex quantity containing a real and an imaginary part,
\begin{eqnarray}
 \phi_s(k,x)= - \int_{x}^{x_f} {1 \over v} V(k,x') dx_\|'.
\end{eqnarray}
As $x_f$ is a function of $x$, the phase function $\phi_s(x\to x_f)$
is a function of $k$ and $x$ and can be abbreviated as $\phi_s(k,x)$.

With this modification, the derivation of the single-particle and
two-particle distributions can be carried out as in the case with the
Glauber theory.  The two-particle correlation function $R(k_1,k_2)$
becomes
\begin{eqnarray}
\label{rkk1}
R(k_1,k_2) = 
\left |  \int
d^4x \, e^{ i(k_1-k_2) \cdot x +i\phi_c(x, k_1 k_2 )}
e^{i  (\phi_s (k_1,x)-\phi_s^* (k_2,x)) }
\rho_{\rm eff} (x; k_1 k_2) \right |^2,
\end{eqnarray}
where
\begin{eqnarray}
{\cal R}e~ (\phi_s (k_1,x)-\phi_s^* (k_2,x))
=-
\int_{x}^{x_d} 
\left ( \frac{1}{v_1} {\cal R}e~ V(k_1,x') dx_\|'
-       \frac{1}{v_2} {\cal R}e~ V(k_2,x') dx_\|'
\right ),
\end{eqnarray}
\begin{eqnarray}
{\cal I}m~ (\phi_s (k_1,x)-\phi_s^* (k_2,x))
&=&- \int_{x}^{x_d} 
\left ( \frac{1}{v_1} {\cal I}m~ V(k_1,x') dx_\|'
+       \frac{1}{v_2} {\cal I}m~ V(k_2,x') dx_\|'
\right )\nonumber\\
&=&2 {\cal I}m~ {\bar \phi}_s (x)
\end{eqnarray}
and $v_i=|(\bbox{k}_i)_\||/k_i^0 $.  For low-energy detected
particles, the velocities of the two detected particles can differ
substantially and the real part of the phase function does not cancel
exactly.  There will be a substantial correction to the two-particle
correlation function at low energies \cite{Chu94,Sho00}.  For detected
pions particles with kinetic energies close to or greater than its
rest mass, in which we shall focus our attention, $v_i\sim 1$.  The
real parts of the phase function $\phi_s(k,x)$ cancel and there will
be negligible modification of the two-particle correlation function.
We again obtain the results of Eqs.\ (\ref{pk1mscat}),
(\ref{pk2mscat}), and (\ref{pk3mscat}) for the single-particle
distribution, and the result of Eq.\ (\ref{2b1mscat}) for the
two-particle distribution.

In applying the Glauber theory at high energies or the optical model
at intermediate energies, the precise energy at which one needs to
switch from the Glauber theory to the optical model needs not concern
us here at present as both descriptions lead to a complex phase shift
function $\phi_s(x)$.  The imaginary part of the phase shift function
gives rise to an absorption.  The real part of the phase shift leads
to negligible modification of the two-particle correlation function.
The conclusion concerning the real phase of coordinate is a rather
general result.  Any final-state interaction, that leads to an
additional real phase function of coordinate and a weak dependence on
momentum, does not modify the two-particle correlation.  As a
consequence, the effective density distribution depends on the initial
source distribution and not necessarily on the freeze-out
distribution.  The intensity interferometry arises from the difference
in the phases when the two particles takes on two different sets of
histories.  For either set of histories depicted in Fig.\ 2, even
though the amplitude for the propagation from $x_1$ to $x_{f1}$
depends on the positions of the medium scatterer which can be random,
the real parts of the phases in propagating from the source point to
the freeze-out points are the same.  They cancel out when we evaluate
the HBT two-particle correlation function $R(k_1,k_2)$.

Previously, Gyulassy $et~al.$ found that the two-particle correlation
function for a coherent source is unaffected by a final-state
mean-field potential. For a chaotic source, which is the object of our
interest here, they expressed the two-particle correlation function in
terms of distorted waves in the mean-field [Eq.\ (5.32) of
\cite{Gyu79}]. The present results for the optical potential
corresponds to an explicit evaluation of this two-particle correlation
function using the eikonal approximation for high-energy particles.
In the process of this evaluation, we find that the phase distortion
due to a real mean-field potential cancels out, as we remarked
earlier, and lead to negligible effect on the two-particle correlation
at high particle energies.  The distortions can however be substantial
for low-energy particles as discussed in \cite{Chu94,Sho00}.

\section{Conclusions and Discussions}

We start from the probability amplitude written in terms of the path
integral over the classical trajectory for a single particle and a
pair of identical particles.  The sum over the probability amplitudes
for a chaotic source leads to an intensity interference for the
detection of two identical particles. 

As the detected particles traverse from the source point to the
freeze-out point, they are subject to scattering with particles in the
medium.  We have examined the effects of multiple scattering.  Because
the Hanbury-Brown-Twiss intensity interferometry is purely a
quantum-mechanical phenomenon, we investigate the problem of multiple
scattering within the quantum-mechanical framework of the Glauber
theory and the optical model.  We find that multiple scattering leads
to an effective density distribution that depends on the initial
source distribution rather than the freeze-out density distribution.
This conclusion follows from the wave nature of the detected particle,
which is an important ingredient for the occurrence of intensity
interferometry.  The Glauber multiple scattering amplitude and the
optical model scattering also represent coherent scattering processes
\cite{Gla59} as the corresponding phase function is an analytical
function of the coordinate. There is the expected absorption arising
from the imaginary part of the phase shift function in the multiple
scattering process.

We have studied the multiple scattering process at high energies using
the Glauber model and at intermediate energies using the optical
model in the eikonal approximation.  At low energies when these simple
approximations may not be valid, it will be of great interest to study
a quantum mechanical description of the multiple scattering process
using either a quantum mechanical many-particle theory or a pion
mean-field optical potential without the eikonal approximation.

In the presence of a collective flow, we find that two-particle
correlation measurements lead to an effective source distribution that
depends on the initial source distribution.  The relevant momentum is
the initial momentum shifted from the detected momentum downward by
the collective flow, as given by Eq.\ (\ref{flow}) obtained here.  In
addition, the collective flow leads to a phase function $\phi_c(x, k_1
k_2)$ in Eq.\ (\ref{gkk}) which will modify the effective two-particle
correlation.  It consists of terms which tend to cancel each other.
We examine sample results of hydrodynamical calculations of Rischke
and Gyulassy \cite{Ris96} for a highly compressed relativistic fluid
initially at rest.  We find that the flow velocity of a fluid element
reaches the asymptotic velocity rapidly at the surface, or in the
interior when the rebounding wave carries the fluid element outward.
There can be a substantial cancellation in the collective flow phase
function $\phi_c(x, k_1 k_2)$.  As a consequence, the effective
density measured in HBT measurements is expected to depend essentially
on the initial source distribution for such an expansion.  Much more
work will be required to examine further the behavior of $\phi_c(x,
k_1 k_2)$ and $R(k_1, k_2)$ for general flows and initial conditions.

While further studies are continuing, it is interesting to explore the
possibility that HBT measurements are indeed related mainly with the
distribution of the initial source distribution.  In that case, we
expect that the initial source transverse dimension should be given
approximately by the spatial dimension of the colliding nuclei which
should be nearly independent of the collision energy. We also expect
that in the initial source prior to the collective flow, the
transverse size in the ``out'' direction should be approximately the
same as that in the ``side'' direction.  Hence, the HBT transverse
radii should be approximately the same as the colliding nuclear radii,
have only a weak collision-energy dependence, and $R_{\rm out}/R_{\rm
side}$ should be approximately close to 1. These expectations are
consistent with the gross features of HBT transverse radii in
high-energy heavy-ion collisions
\cite{Phe02,Sta01,Lis00,Ahl02,Ahl00,Agg00}.

\begin{acknowledgments}
The author would like to thank Prof. R. Glauber for collaborative
discussions on the multiple scattering model and diffractive
transport.  The author wishes to thank Drs.\ T. Awes, V. Cianciolo,
M. Gyulassy, C. Pajares, S. Pratt, S. Sorensen, G. Young, and Weining
Zhang for valuable discussions and comments. The author would like to
thank specially Drs.\ T. Awes and V. Cianciolo for reading through the
manuscript and making insightful suggestions.  The author is indebted
to Professor J. A. Wheeler for his earlier favorable comments on the
treatment of intensity interferometry in author's book
\cite{Won94}.  Professor Wheeler's comments provide additional impetus
to investigate the present problem.  This research was supported by
the Division of Nuclear Physics, Department of Energy, under Contract
No. DE-AC05-00OR22725 managed by UT-Battelle, LLC.
\end{acknowledgments}

\end{document}